\documentclass[aps]{revtex4}
\usepackage{epsfig}
\usepackage{amsmath,amssymb}
\usepackage[english]{babel}
\usepackage[dvips]{color}

\parskip=\medskipamount

\newcommand{\eq}[1]{(\ref{#1})}
\newcommand{\fig}[1]{Fig.\ref{#1}}

\newcommand{\be}{\begin{equation}}
\newcommand{\ee}{\end{equation}}

\newcommand{\barr}{\begin{array}}
\newcommand{\earr}{\end{array}}

\newcommand{\beqn}{\begin{eqnarray}}
\newcommand{\eeqn}{\end{eqnarray}}

\newcommand{\bs}{\begin{subequations}}
\newcommand{\es}{\end{subequations}}

\newcommand\disp{\displaystyle}

\newcommand{\la}{\left<}
\newcommand{\ra}{\right>}

\begin{document}

\title{Unzipping of two random heteropolymers: Ground state energy and finite size effects}

\author{M.V. Tamm$^1$ and S.K. Nechaev$^2$\footnote{Also at: P.N. Lebedev Physical Institute
of the Russian Academy of Sciences, 119991, Moscow, Russia}}

\affiliation{$^1$Physics Department, Moscow State University 119992 Moscow, Russia \\
$^2$LPTMS, Universit\'e Paris Sud, 91405 Orsay Cedex, France}

\date{\today}

\begin{abstract}

We have analyzed the dependence of average ground state energy per
monomer, $e$, of the complex of two random heteropolymers with
quenched sequences, on chain length, $n$, in the ensemble of
chains with uniform distribution of primary sequences. Every chain
monomer is randomly and independently chosen with the uniform
probability distribution $p=1/c$ from a set of $c$ different types
A, B, C, D, .... Monomers of the first chain could form saturating
reversible bonds with monomers of the second chain. The bonds
between similar monomer types (like A--A, B--B, C--C, etc.) have
the attraction energy $u$, while the bonds between different
monomer types (like A--B, A--D, B--D, etc.) have the attraction
energy $v$. The main attention is paid to the computation of the
normalized free energy $e(n)$ for intermediate chain lengths, $n$,
and different ratios $a=\frac{v}{u}$ at sufficiently low
temperatures when the entropic contribution of the loop formation
is negligible compared to direct energetic interactions between
chain monomers and the partition function of the chains is
dominated by the ground state. The performed analysis allows one
to derive the force, $f$, which is necessary to apply for
unzipping of two random heteropolymer chains of equal lengths
whose ends are separated by the distance $x$, averaged over all
equally distributed primary structures at low temperatures for
fixed values $a$ and $c$.

\bigskip


\end{abstract}

\maketitle

\section{Introduction}
\label{sect:1}

Recent progress in nanotechnology has offered a possibility of single--molecular experiments. The
corresponding technique allows one to investigate many physico--chemical and biological properties
of individual molecules. One of the modern biophysical key experiments deals with the mechanical
unzipping of individual double--stranded DNA macromolecule under the action of external force
applied to the ends of strands. This question has been analyzed theoretically in a number of
important contributions \cite{nel_lub1,bhat1,seb,mon,bhat2,nel_lub2,cule,tang,singh}. Some of them
are devoted to the consideration of unzipping transition in an effective homopolymer chain, the
other pay attention to the heterogeneity of primary sequence of complimentary strands constituting
the DNA molecule.

In our work we address a problem of unzipping of a complex of
two random heteropolymers of finite lengths at sufficiently low
temperatures when the partition function is dominated by the
ground state. We demonstrate that this problem can be
mapped to the problem of alignment of two random sequences with
the general "cost function" which takes into account the weights
of perfect matches, mismatches and gaps (all necessary definitions
are introduced below). Using this bijection we are able to compute
the external work necessary to unzip the complex of two random
heteropolymers, averaged over the uniform distribution of all
possible primary sequences of heteropolymers. Our consideration
allows also to conjecture the scaling corrections to the leading
behavior of the force fluctuations due to the finiteness of the
lengths of heteropolymer chains.

The paper is organized as follows. In Section \ref{sect:2} we
define a model under consideration and introduce the basic
notations. In Section \ref{sect:3} we consider unzipping of two
random heteropolymers from the point of view of the search of
Longest Common Subsequence (LcS) of two random sequences. The
expectation of the LCS energy is considered in Section
\ref{sect:4}. In Conclusion we give the qualitative explanation of
our main results and derive a force, which is necessary to apply
to the chain ends to unzip two random heteropolymer chains at low
temperatures.

\section{The model}
\label{sect:2}

\begin{figure}[ht]
\epsfig{file=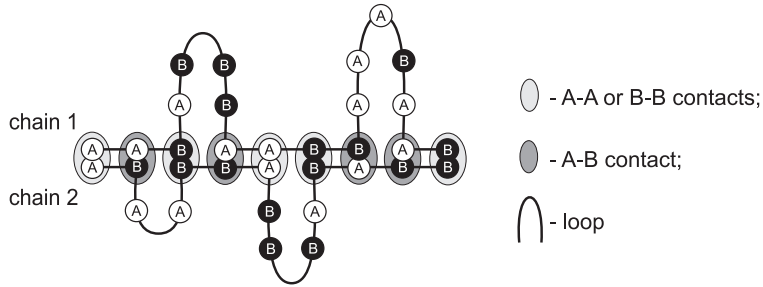,width=10cm} \caption{Schematic picture of a complex of two random
heteropolymer chains.} \label{fig:1}
\end{figure}

Consider two random heteropolymer chains of lengths $L_1=m\ell$
and $L_2=n\ell$ correspondingly. In what follows we shall measure
the lengths of the chains in number of monomers, $m$ and $n$,
supposing that the size of an elementary unit, $\ell$, is equal to
1. Every monomer can be randomly and independently chosen with the
uniform probability distribution $p=\frac{1}{c}$ from a set of $c$
different types A, B, C, D, ... . Monomers of the first chain
could form saturating reversible bonds with monomers of the second
chain. The term "saturating" means that any monomer can form a
bond with at most one monomer of the other chain. The bonds
between similar types (like A--A, B--B, C--C, etc.) have the
attraction energy $u$ and are called below "matches", while the
bonds between different types (like A--B, A--D, B--D, etc.) have
the attraction energy $v$ and are called "mismatches". Some parts
of the chains could form loops hence contributing to the entropic
part of the free energy of the system. Schematically a particular
configuration of the system under consideration for $c=2$ is shown
in \fig{fig:1}.

Our aim is to compute the free energy of the described model at sufficiently low temperatures when
the entropic contribution of the loop formation is negligible compared to the energetic part of the
direct interactions between chain monomers.

\begin{figure}[ht]
\epsfig{file=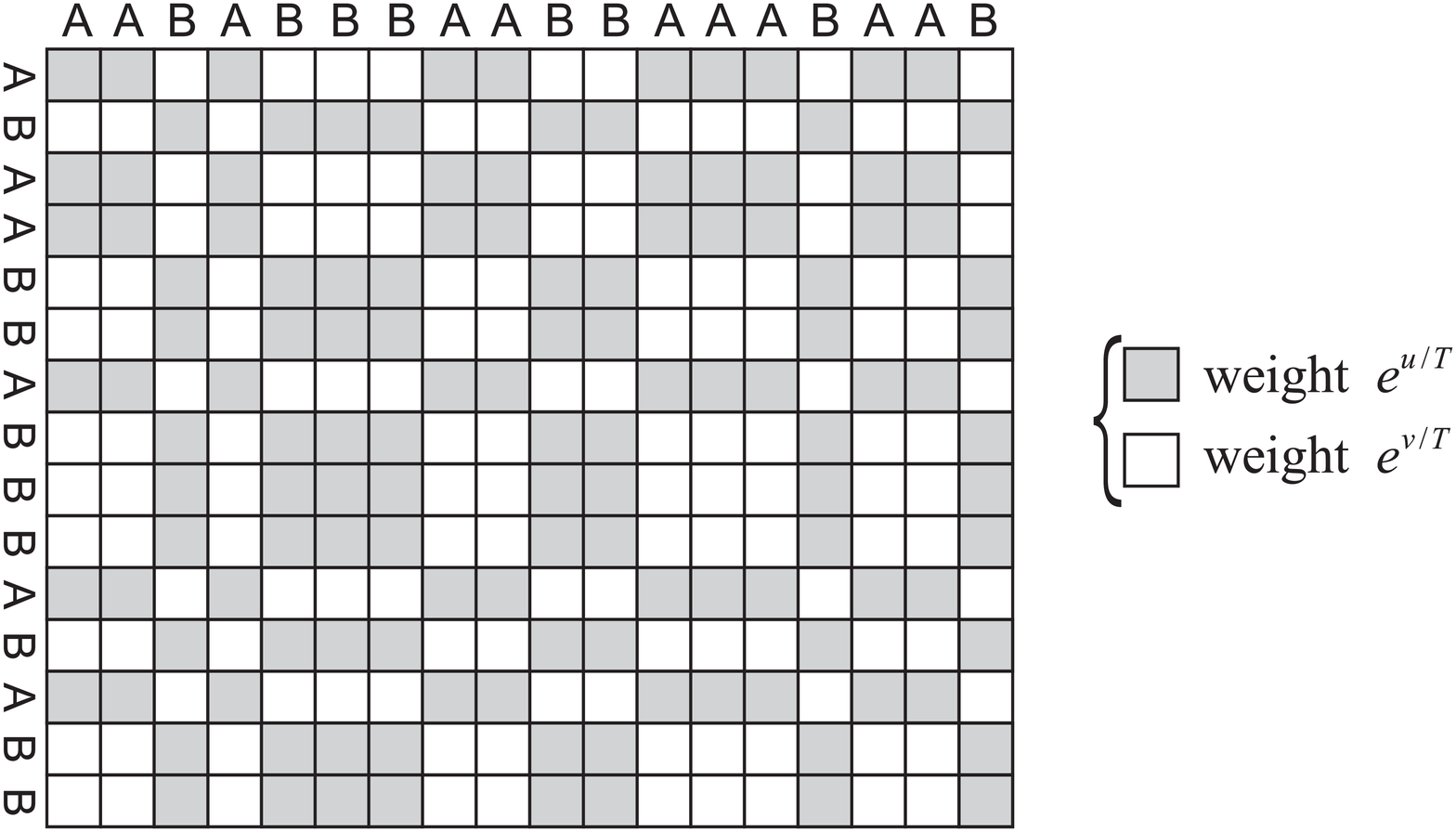,width=10cm} \caption{Contact map $\{\beta\}$ corresponding to the complex
of two random heteropolymer chains shown in \fig{fig:1}.}
\label{fig:1c}
\end{figure}

Consider now the partition function of such a complex $G_{m,n}$ which is
the sum over all possible arrangements of bonds. Since we are interested in the
low--temperature behavior of $G_{m,n}$, we neglect the entropic contribution of the loop weights
which allows to write $G_{m,n}$ recursively in terms of the partition functions of individual chains
$g(n)$:
\be
\left\{\barr{l} \disp G_{m,n}=g(m) g(n) +\sum_{i,j=1}^{m,n}\beta_{i,j}\, G_{i-1,j-1} g(m-i) g(n-j)\medskip \\
G_{m,0}=g(m);\;\; G_{0,n}=g(n);\;\; G_{0,0}=1 \earr \right.
\label{eq:1} \ee The meaning of the equation \eq{eq:1} is as
follows. Starting from, say, the left ends of the chains shown in
\fig{fig:1} we find the first actually existing contact between
the monomers $i$ (of the first chain) and $j$ (of the second
chain) and sum over all possible arrangements of this first
contact. The first term ''1'' in \eq{eq:1} means that we have not
found any contact at all. The entries $\beta_{i,j}$ ($1\le i\le m,
\; 1\le j\le n$) are the statistical weights of the bonds which
are encoded in a contact map $\{\beta\}$: \be
\beta_{m,n} = \begin{cases} \beta^+\equiv e^{u/T} & \mbox{if monomers $i$ and $j$ match} \\
\beta^-\equiv e^{v/T} & \mbox{if monomers $i$ and $j$ do not match}
\end{cases}
\label{eq:2}
\ee
For a system of two heteropolymer chains depicted in \fig{fig:1} the contact map $\{\beta\}$ is
shown in \fig{fig:1c}.

Generally speaking, if one allows for loop formation within a single chain, the
partition functions of individual chains $g(n)$ satisfy, in turn, recurrent
equation \cite{de Gennes, Erukh, Mueller}
\be
g(n)=1+\sum_{i=1}^{n-1}\sum_{j=i+1}^{n}\beta'_{i,j}g(j-i-1)g(n-j);\;\; g(0)=1
\label{eq:self}
\ee
where $\beta'_{i,j}$ are the constants of self-association, which are, similarly to $\beta_{m,n}$,
random variables encoded by some contact map. However, for the sake of simplicity, we assume in this paper
that there is no self-association in the systems under discussion, and thus
\be
\beta'\equiv 0;\;\; g(n) \equiv 1
\label{eq:self0}
\ee

This simplification is, in fact, rather significant from the mathematical point of view
since we replace the quadratic set of equations \eq{eq:1}, \eq{eq:self} with the
linear set \eq{eq:1}, \eq{eq:self}. However, we expect the intrachain association
to be effectively suppressed due to the finite flexibility of the single chains, and,
therefore, the typical values of $\beta'$ to be much less then the typical values of
$\beta$. We expect thus the interchain association to give just some minor corrections to
the results obtained below.

The case of non-zero $\beta'$s was thoroughly investigated recently \cite{Bund2, Neher, TN2}
in a set-up when the interactions in the system are predetermined instead of random
and we refer the reader to these papers for more detail.

\section{Unzipping of two random heteropolymers and search of longest common subsequence (LCS) of two
random sequences}
\label{sect:3}

\subsection{Heteropolymer ground state energy: local recursive construction}
\label{sect:3a}

The straightforward computation shows that the partition function $G_{m,n}$ obeys the following
exact {\em local} recursion
\be
G_{m,n} = G_{m-1,n} + G_{m,n-1} + (\beta_{m,n}-1)\, G_{m-1,n-1}
\label{eq:3}
\ee
Note that if $\beta_{i,j}=2$ for all $1\le i\le m$ and $1\le j\le n$, the recursion relation
\eq{eq:3} generates the so-called Delannoy numbers \cite{delannoy}.

Represent now the partition function $G_{m,n}$ in the following way
\be
G_{m,n}=e^{F_{m,n}/T}
\label{eq:4}
\ee
where $-F_{n,m}$ has the sense of the free energy and $T$ stands for the temperature of the complex
of two heterogeneous chains of lengths $m$ and $n$. Considering the $T\to 0$ limit, we get
\be
F_{m,n} = \lim_{T\to 0} T \ln \Big( e^{F_{m-1,n}/T} + e^{F_{m,n-1}/T} + (\beta_{m,n}-1) \,
e^{F_{m-1,n-1}/T} \Big)
\label{eq:5}
\ee
which can be regarded as the equation for the ground state energy of a chain. The expression
\eq{eq:5} can be rewritten in a symbolic form
\be
F_{m,n} = \max \left[F_{m-1,n},\, F_{m,n-1},\, F_{m-1,n-1} + \eta_{m,n} \right]
\label{eq:6a}
\ee
where
\be
\eta_{m,n} = T\ln (\beta_{m,n}-1)= \begin{cases} \eta^+ = T\ln (e^{u/T} -1)  & \mbox{in case of
match} \\ \eta^- = T\ln (e^{v/T} -1) & \mbox{in case of mismatch} \end{cases}
\label{eta}
\ee
Taking $\eta^+$ as the unit of the energy, we can rewrite \eq{eq:6a} as follows
\be
\tilde{F}_{m,n} = \max \left[\tilde{F}_{m-1,n},\, \tilde{F}_{m,n-1},\, \tilde{F}_{m-1,n-1} +
\tilde{\eta}_{m,n} \right]
\label{eq:6aa}
\ee
where
\be
\tilde{\eta}_{m,n} = \begin{cases} 1 & \mbox{in case of match} \\ \disp a = \frac{\eta^-}{\eta^+} &
\mbox{in case of mismatch} \end{cases}
\label{tilde_eta}
\ee
In the low--temperature limit the parameter $a$ has simple expression in terms of coupling
constants $u$ and $v$:
\be
a= \frac{\eta^-}{\eta^+} = \left.\frac{\ln(e^{v/T} -1)}{\ln(e^{u/T} -1)}\right|_{T\to 0} =
\frac{v}{u}
\label{eq:15}
\ee
Finally, the initial conditions for $\tilde{F}_{m,n}$ transform due to the second of equations
\eq{eq:1} into
\be
\tilde{F}_{0,n}=\tilde{F}_{n,0}=\tilde{F}_{0,0}=0
\label{eq:6b}
\ee

\subsection{Matching with gaps: the cost function}
\label{sect:3b}

In Eqs.\eq{eq:6a}--\eq{eq:6b} we can recognize the recursive algorithm \cite{Gusfield,Monvel} for
the determination of the length $F_{m,n}$ of the Longest Common Subsequence (LCS) of two arbitrary
sequences of lengths $m$ and $n$. It is easy to see that the search of $F_{m,n}$ can be completed
in polynomial time $\sim O(mn)$.

Recall that the problem of finding the LCS in a pair of sequences drawn from alphabet of $c$
letters is formulated as follows. Consider two sequences $\alpha=\{\alpha_1, \alpha_2,\dots,
\alpha_m\}$ (of length $m$) and $\beta=\{\beta_1, \beta_2,\dots, \beta_n\}$ (of length $n$). For
example, let $\alpha$ and $\beta$ be two random sequences of $c=4$ base pairs A, C, G, T of a DNA
molecule, e.g., $\alpha=\{\rm A, C, G, C, T, A, C\}$ with $m=6$ and $\beta=\{\rm C, T, G, A, C\}$
with $n=5$. Any subsequence of $\alpha$ (or $\beta$) is an ordered sublist of $\alpha$ (and of
$\beta$) entries which need not to be consecutive, e.g, it could be $\{\rm C, G, T, C\}$, but not
$\{\rm T, G, C\}$. A common subsequence of two sequences $\alpha$ and $\beta$ is a subsequence of
both of them. For example, the subsequence $\{\rm C, G, A, C\}$ is a common subsequence of both
$\alpha$ and $\beta$. There are many possible common subsequences of a pair of initial sequences.
The aim of the LCS problem is to find the longest of them. This problem and its variants have been
widely studied in biology \cite{NW,SW,WGA,AGMML}, computer science \cite{SK,AG,WF,Gusfield},
probability theory \cite{CS,Deken,Steele,DP,Alex,KLM} and more recently in statistical physics
\cite{ZM,Hwa,Monvel}. A particularly important application of the LCS problem is to quantify the
closeness between two DNA sequences. In evolutionary biology, the genes responsible for building
specific proteins evolve with time and by finding the LCS of similar genes in different species,
one can learn what has been conserved in time. Also, when a new DNA molecule is sequenced {\it in
vitro}, it is important to know whether it is really new or it is similar to already existing
molecules. This is achieved quantitatively by measuring the LCS of the new molecule with other ones
available from database.

\begin{figure}[ht]
\epsfig{file=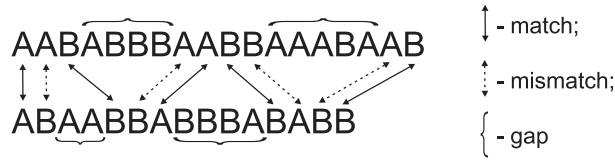,width=8cm} \caption{Matches, mismatches and gaps in a pair of sequences
corresponding to the configuration of two random heteropolymers shown in \fig{fig:1}.}
\label{fig:2}
\end{figure}

In the simplest version of the LCS problem only the number of perfect matches is taken into
account, i.e. there is no difference between mismatches and gaps. One can, however, easily
construct a generalized model where this difference comes into play. Let us introduce the general
"cost function", ${\cal S}$, having a meaning of an energy (see, for example \cite{Hwa97,HWA2} for
details)
\be
{\cal S} = N_{\rm match} + \mu\, N_{\rm mis} + \delta\, N_{\rm gap}
\label{eq:7}
\ee
In \eq{eq:7} $N_{\rm match}$, $N_{\rm mis}$ and $N_{\rm gap}$ are correspondingly the numbers of
matches, mismatches and gaps in a given pair of sequences---see \fig{fig:2}, and $\mu$ and $\delta$
are respectively the energies of mismatches and gaps. Without the loss of generality the energy of
matches can be always set to 1. Besides \eq{eq:7} we have an obvious conservation law
\be
n + m = 2 N_{\rm match} + 2 N_{\rm mis} + N_{\rm gap}
\label{eq:8}
\ee
which allows one to exclude $N_{\rm gap}$ from \eq{eq:7} and rewrite this expression as follows:
\be
{\cal S} = N_{\rm match} + \mu N_{\rm mis} + \delta (n + m - 2 N_{\rm match} - 2 N_{\rm mis}) = (1
- 2\delta) N_{\rm match} + (\mu - 2\delta) N_{\rm mis} + {\rm const}
\label{eq:9}
\ee
In \eq{eq:9} the irrelevant constant $\delta (n+m)$ can be dropped out.

Now we can adopt $(1 - 2\delta)$ as a unit of energy. Finally we arrive at the following expression
\be
\tilde{\cal S} = N_{\rm match} + \gamma N_{\rm mis}
\label{eq:10}
\ee
where
\be
\gamma=\frac{\mu - 2 \delta}{1 - 2 \delta},
\label{eq:11}
\ee
and $\gamma \leq 1$ by definition. The interesting region is $0 \le \gamma \le 1$, since otherwise
there are no mismatches at all in the ground state (i.e., there is no difference between
$\gamma=0$, which corresponds to simplest version of the LCS problem, and $\gamma<0$).

It is known \cite{Hwa97,HWA2} that the ground state energy
\be
\tilde{\cal S}^{\rm max}= \max\left[N_{\rm match} + \gamma N_{\rm mis}\right]
\label{eq:12}
\ee
satisfies the recursion relation
\be
\tilde{\cal S}^{\rm max}_{m,n} = \max \left[\tilde{\cal S}^{\rm max}_{m-1,n},\, \tilde{\cal S}^{\rm
max}_{m,n-1},\, \tilde{\cal S}^{\rm max}_{m-1,n-1} + \zeta_{m,n} \right]
\label{eq:12a}
\ee
with
\be
\zeta_{m,n}= \begin{cases} 1 & \mbox{in case of match} \\ \gamma & \mbox{in case of mismatch}
\end{cases}
\label{eq:13}
\ee
Indeed, the ground state may correspond either (i) to the last two monomers connected, then the
ground state energy equals $\tilde{\cal S}^{\rm max}_{m-1,n-1}+\zeta_{M,N}$, or (ii) to the
unconnected end monomer of the fist (or second) chain, then the ground state energy is $\tilde{\cal
S}^{\rm max}_{m,n-1}$ (or $\tilde{\cal S}^{\rm max}_{m-1,n}$).

Comparing Eqs\eq{eq:12a}, \eq{eq:13} with Eqs.\eq{eq:6aa}, \eq{tilde_eta} one sees that they are
identical up to the exchange of variables $\gamma \leftrightarrow a$. This establishes the analogy
between initial heteropolymer problem formulated in \eq{eq:1}--\eq{eq:2} in the low--temperature
limit \eq{eq:6aa} and the standard matching problem with general cost function \eq{eq:7}.

For a pair of fixed sequences of lengths $m$ and $n$, the cost function $\tilde{\cal S}^{\rm
max}_{m,n}$ is just a number. In the stochastic version of the LCS problem one compares two random
sequences drawn from alphabet of $c$ letters and hence the cost function $\tilde{\cal S}^{\rm
max}_{m,n}$ is a random variable. We are interested in the computation of the expectation and the
variance of $\tilde{\cal S}^{\rm max}_{m,n}$ for $m=n\gg 1$ and the interpretation of the obtained
results for LCS in terms of initial problem of unzipping of two random heteropolymers.

\subsection{Bernoulli model for heteropolymers}
\label{sect:3c}

We should note that the variables $\tilde{\eta}_{m,n}$ in \eq{eq:6a} are not independent of each
other. Actually, consider a simple example of two strings $\alpha={\rm AB}$ and $\beta={\rm AA}$.
One has by definition: $\tilde{\eta}_{1,1}=\tilde{\eta}_{1,2}=1$ and $\tilde{\eta}_{2,1}=0$. The
knowledge of these three variables is sufficient to predict that the last two letters do not match
each other, i.e., $\tilde{\eta}_{2,2}=0$. Thus, $\tilde{\eta}_{2,2}$ can not take its value
independently of $\tilde{\eta}_{1,1},\, \tilde{\eta}_{1,2},\, \tilde{\eta}_{2,1}$. These residual
correlations between the $\tilde{\eta}_{i,j}$ variables make the LCS problem very complicated.
However for two random sequences drawn from the alphabet of $c$ letters, the correlations between
the $\tilde{\eta}_{m,n}$ variables vanish for $c\to \infty$.

In our work we restrict ourselves with the so-called Bernoulli matching (BM) model \cite{Monvel}
(which is simpler but yet nontrivial variant of the original LCS problem) where one ignores the
correlations between $\tilde{\eta}_{m,n}$ for all $c$. The cost function $\tilde{F}_{m,n}^{BM}$ of
the BM model satisfies the same recursion relation \eq{eq:6a} except that the
$\tilde{\eta}_{m,n}$'s are now independent variables, each drawn from the bimodal distribution:
\be
\tilde{\eta}=\begin{cases} a & \mbox{with probability $P(\tilde{\eta})=1- \frac{1}{c}$} \medskip \\
1 & \mbox{with probability $P(\tilde{\eta})=\frac{1}{c}$} \end{cases}
\label{eq:16}
\ee
As it has been already said, this approximation is expected to be exact only in the appropriately
taken $c\to \infty$ limit. Nevertheless, for finite $c$, the results on the BM model can serve as a
useful benchmark for original LCS model to decide if indeed the correlations between
$\tilde{\eta}_{m,n}$ are important or not.

Note that the problem under discussion can be redefined as follows. Consider a matrix $\tilde\eta$
of size $m\times n$ and let the elements of this matrix be independent random variables with
bimodal distribution (\ref{eq:16}). Consider now all directed paths in this matrix, i.e. ordered
sequences $\{(m_1,n_1); (m_2,n_2);...; (m_k,n_k)\}$ such that $m_i>m_{i-1}$ and $n_i>n_{i-1}$ for
$i=2,...,k$. Calculating the ground state energy of the matching problem is obviously equivalent to
maximizing the sum of the matrix elements along these directed trajectories:
\be
E_{m,n}(a)=\max_{\text{all sequences}} \sum_{i=0}^k \tilde\eta_{m_i,n_i}
\label{eq:18}
\ee

\begin{figure}[ht]
\epsfig{file=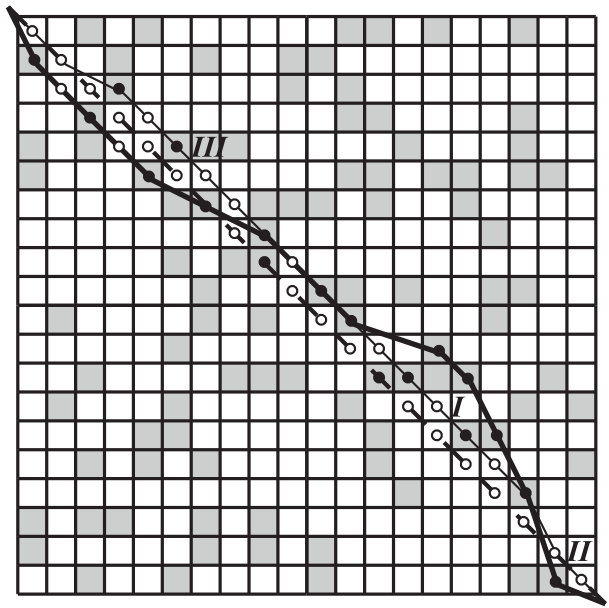,width=8cm} \caption{An example of a random distribution of "1"s (gray
squares) and "$a$"s (white squares) on a $20\times 20$ matrix with $c = 4$. The optimal path for
$a=0$ is shown by the thick line, the diagonal optimal path for $a=1$ -- by the dashed line and the
evolution of the optimal path with increase of $a$ -- by thin line. The "1"s and "$a$"s lying on
the optimal paths are additionally marked by filled and open circles, respectively. See the main
text for more details.}
\label{fig:3}
\end{figure}

In \fig{fig:3} we show an example of the evolution of the optimal path with the increase of $a$ for
some particular random distribution of weights "a" and "1" (shown by white and grey squares
respectively) corresponding to $c=4$.

The optimal path for small $a$ is drawn in bold in \fig{fig:3}. With the increase of $a$, the first
change in the optimal path configuration happens at $a=\frac{1}{3}$ when a shortcut I (shown by a
thin line) is formed instead of the corresponding section of the bold line. Then, at
$a=\frac{1}{2}$ the shortcut marked by II actuates, then at $a=\frac{2}{3}$ the one marked by III
comes into play. So, for $a>\frac{2}{3}$ the optimal path is III--I--II. In what follows we call
this kind of path {\it subdiogonal}, meaning that it goes only through the diagonal of the matrix
($a_{i,i}$ for $i=1,...,n$) and one of its subdiagonals ($a_{i,i+1}$, or $a_{i+1,i}$ for
$i=1,...,n-1$). Finally, at $a=\frac{5}{6}$ the subdiagonal path III--I--II ceases to be the
optimal one, and optimal path sticks to the diagonal (dashed line) where it stays up to $a=1$.

\section{Expectations of LCS energy for general cost function $\tilde{\cal S}$}
\label{sect:4}

In this Section we consider the dependence of the ground state energy on the parameter $a$ defined
in Eqs.\eq{eq:13}--\eq{eq:15}. We start with the consideration of the limiting cases: (i) $a \ll 1$
and (ii) $\epsilon=1-a \ll 1$ and then, with the physical insight in hands, proceed to the
semi--quantitative consideration of the general case.

\subsection{The case $0<a=\frac{u}{v}\ll 1$}
\label{sect:4a}

In the limit $a=0$, as we have mentioned before, the problem under consideration corresponds
exactly to the simplest version of the Longest Increasing Subsequence (LCS) problem, where the
mismatches have no cost at all. The Bernoulli Matching model for this problem has been considered
in details in \cite{nech_maj}. An example of the random matrix with the optimal path is outlined by
the bold line in \fig{fig:3} (only filled circles, i.e. points with the weight equal to 1 are
relevant in this case). We know that the ground state energy, $E_{m,n}$, as a function of the chain
lengths $m,n$ behaves asymptotically for large $m$ and $n$ as
\be
E_{m,n}(c,a=0) = \frac{2\sqrt{p m n}- p(m+n)}{q} + \frac{(p m n)^{1/6}}{q} \left[(1+p)-
\sqrt{\frac{p}{m n}}(m+n)\right]^{2/3} \chi
\label{nech_maj_energy}
\ee
where $p=c^{-1}$, $q=1-p$ and $\chi$ is a random variable with the Tracy--Widom distribution
\cite{TW}. The ground state energy, $E_{m,n}(a=0)$, has a meaning of the LIS length of "1" (see
\cite{nech_maj}). The mean value $\la E_{m,n} \ra$ in the thermodynamic limit $n=m\to \infty$
equals to
\be
\la E_{m,n} \ra \equiv \la E_{n,n} \ra  = 2\frac{\sqrt{p}-p}{q}n = \frac{2}{1+\sqrt{c}}n
\label{nech_maj_2}
\ee

Consider now the case of finite $a=\frac{u}{v}$ paying special
attention to the effects of finite values of $m,n$ on typical
fluctuations of $E$. We assume below $m=n$ for simplicity.

If the value of $a$ is small but finite ($0<a=\frac{u}{v}\ll 1$, the meaning of "small" is
specified below), then the trajectory of the optimal matching path does not change with respect to
the case of $a=0$. The only difference from the $a=0$ case is that there are mismatches inserted
between the matches whenever it is possible (see open circles along the bold line in \fig{fig:3}).
It is not difficult to estimate the number of such inserted mismatches. Namely, the typical
distance $\la d\ra$ between the consequent "1" (i.e. gray squares) along the optimal path in
\fig{fig:3} projected to the horizontal and vertical axes is, correspondingly, $\la m_{i+1}-m_i\ra$
and $\la n_{i+1}-n_i\ra$. The value of $\la d\ra$ is dictated by the density of black circles along
the optimal path (see fig.\fig{fig:3}). For $m=n \to \infty$ one has
\be
\la d\ra=\langle m_{i+1}-m_i\rangle=\langle n_{i+1}-n_i\rangle =\frac{n}{\la E_{n,n} \ra} =
\frac{1+\sqrt{c}}{2} \label{0_dist}
\ee
The average energy gain due to $a$'s (i.e. white squares in \fig{fig:3}) inserted into the optimal
path can be estimated as follows
\be
\la \Delta E\ra = \la E_{n,n} \ra \Big(\la \min [m_{i+1}-m_i,\; n_{i+1}-n_i] \ra - 1 \Big) a
\label{0_deltaE}
\ee
Indeed, we can insert a white square into the optimal path between consequent gray squares if and
only if the distance between these consequent gray squares in each of the dimensions is bigger or
equal than two (we measure the distance in elementary squares). Let us estimate $\la \Delta E\ra$
from above and from below.

1. The upper bound corresponds to the assumption that the increments of $m$ and $n$ are fully
correlated. In this case $\la \min [m_{i+1}-m_i,\; n_{i+1}-n_i] \ra=\la d \ra$ with $\la d \ra$
computed in \eq{0_dist}. Therefore, for $\la \Delta E \ra$ we obtain the following estimate
\be
\la \Delta E \ra< \la E_{n,n}\ra \Big(\la d \ra - 1 \Big) a = \left(1-\frac{2}{1+\sqrt{c}}\right)
na
\label{delta_up}
\ee

2. The construction of the lower bound corresponds to the assumption that the increments of $m$ and
$n$ are completely independent. The computations in this case are slightly more involved since we
have to compute explicitly the average value of the minimum $d_{\min}$ of two independent
increments $m$ and $n$. The computations presented in the Appendix \ref{app:a} lead us to the
following lower bound of $\la \Delta E \ra$:
\be
\la \Delta E \ra>  \la E_{n,n}\ra \Big(\la d_{\min} \ra - 1 \Big)a =
\left(\frac{1+\sqrt{c}}{2\sqrt{c}}-\frac{2}{1+\sqrt{c}}\right) na
\label{E_min1}
\ee

Collecting \eq{delta_up} and \eq{E_min1} we arrive at the following bilateral estimate of $\la
\Delta E\ra $ for $0<a\ll 1$:
\be
\left(\frac{1+\sqrt{c}}{2\sqrt{c}}-\frac{2}{1+\sqrt{c}}\right)a < \frac{\la \Delta E \ra}{n} <
\left(1-\frac{2}{1+\sqrt{c}}\right)a
\label{E_ineq}
\ee

It is worthwhile to notice in advance that, according to the numerical simulations, the genuine values of
$\la \Delta E \ra / n$ are actually very close to the lower bound \eq{E_min1}.

\subsection{The case $a=1-\epsilon$ ($0<\epsilon\ll 1$)}
\label{sect:4b}

Turn now to the opposite situation, $a=1-\epsilon$ ($0<\epsilon\ll 1$). For $\epsilon=0$ the
situation is trivial. Indeed, there is no difference between "1"s and "$a$"s (i.e., gray and white
squares at \fig{fig:3} are identical) and the optimal path is thus the diagonal one with the energy
\be
E(m,n)\equiv \min[m,n]; \;\; E(n,n) \equiv n
\label{a1}
\ee

Now, for small but finite $\epsilon$ and not too long trajectories, $n$ (the definition of "not too
long" is, once again, to be given below), the longest possible path still sticks to the main
diagonal (see \fig{fig:3}). This path is optimal with the ground state energy given by
\be
E^{\rm diag}_n(a) = n - k \epsilon
\label{E_diag}
\ee
where $k$ is the number of $a$'s on the diagonal, which is a random variable distributed with the
binomial law
\be
W(k,n)=\frac{n!}{k!(n-k)!}q^k p^{n-k}
\label{binom}
\ee
(recall that $q=1-\frac{1}{c}$ and $p=\frac{1}{c}$). Hence the average energy $\la E^{\rm
diag}_n\ra$ per monomer on the diagonal path equals
\be
\frac{1}{n}\la E^{\rm diag}_n \ra = 1- \frac{\la k \ra}{n}\, \epsilon = 1-\left(1-p \right)\epsilon
\label{E_diag2}
\ee

Let us estimate now the length, $n_{\rm d}$, on which the optimal path detaches from the main
diagonal. The optimal path of length $n$ is separated from each of the {\em suboptimal} ones
(i.e., those of length $n-1$) by the energy gap $\delta E$:
\be
\delta E = (n - k \epsilon) - (n-1- k' \epsilon) = 1-\epsilon\, \delta k
\label{gap}
\ee
where $\delta k=k-k'$ is the difference in the number of $a$'s on the optimal (diagonal) path and
on the best of the suboptimal paths of lengths $n-1$ (see \fig{fig:3}). The optimal path detaches
from the diagonal when $\delta E<0$. Since $\delta k$ cannot exceed $n-1$, the diagonal path is
always optimal until
\be
1- \epsilon\, \delta k <0 \quad \Rightarrow \quad 1-(n_{\rm d}-1)\epsilon<0
\quad \Rightarrow \quad n_{\rm d}>\epsilon^{-1}+1
\label{eq_a}
\ee
where $n_{\rm d}$ is the length of the optimal path which detaches from the diagonal at energy
$\epsilon$. The inequality \eq{eq_a} gives rather crude lower bound for the value of $n$ for which
the detachment of the optimal path from the diagonal actually happens. To acquire better bounds we
should take into account the concurrent effects involved. On one hand, the {\it single} diagonal
path has the advantage of being the longest one. The corresponding value of $k$ has a binomial
distribution \eq{binom} with the mean $\la k \ra = n q$. On the other hand, the suboptimal paths
(i.e., those of lengths $n-1$) are disadvantageous because they are shorter, however their
intrinsic advantage consists in high degeneracy: one has {\it many} such suboptimal trajectories.
The number $k'$ of $a$'s on each particular suboptimal path is a binomial distributed random
variable with the probability density $W(k',n-1)$ and the mean $\la k' \ra = (n-1) q$.
Now we have to find the {\it best} (i.e. the minimal) value $\la k' \ra $ among ${\cal N}$
suboptimal paths. These suboptimal paths (there are ${\cal N}\sim n^2/2$ of them) are, however, not
independent. It is easy to understand that the number of independent suboptimal paths, $N_{\rm
ind}$, satisfies the following bilateral inequality:
\be
2 \leq N_{\rm ind} \leq 3n-2
\label{ind}
\ee
Indeed, on one hand, there are at least 2 independent paths coinciding with upper and lower
subdiagonals. On the other hand, by definition, the suboptimal paths can visit only these two
subdiagonals and the main diagonal itself. The corresponding energetic costs are therefore always
linear combinations of the values on the diagonal ($n$) and two subdiagonals ($(n-1)$), that is,
$n+2(n-1)=3n-2$ accessible matrix elements, which are themselves independent random variables.
Evidently one cannot construct more than $3n-2$ independent linear combinations out of $3n-2$
independent variables. We are, hence, to compute the {\it average minimum}
of $N_{\rm ind}$ independent random quantities each distributed with the probability density
$W(k',n-1)$. This task is solved in Appendix \ref{app:b}. Taking into account the inequality
\eq{ind} which defines the boundaries of $N_{\rm ind}$, we can get the upper and lower estimates
for $\la \delta k_{N_{\rm ind}} \ra$ ($n \gg 1$), where $\la \delta k_{N_{\rm ind}} \ra$ is defined
as follows:
\be
\la \delta k_{N_{\rm ind}} \ra = \la k\ra -\la k'_{N_{\rm ind}} \ra \equiv n p q - \la k'_{N_{\rm
ind}} \ra
\label{k_ind}
\ee
Substituting into \eq{k_ind} the expressions derived in Appendix \ref{app:b} for $\la k'_{N_{\rm
ind}} \ra$, we have:
\be
q + \frac{1}{\sqrt{\pi}}(npq)^{1/2} <\la \delta k_{N_{\rm ind}} \ra < q + (2n p q)^{1/2} \left[\ln
\left(3 n^{3/2} (p q)^{1/2}\right) \right]^{1/2}
\label{deltaK}
\ee
Remembering now that the optimal path detaches from the diagonal at $\la \delta k_{N_{\rm ind}} \ra
\sim \epsilon^{-1}$, and dropping out all constants of order of one, we arrive for $n \gg1$ at the
following approximate bilateral estimate for the detachment length, $n_{\rm d}$:
\be
n_{\rm d} \lesssim \left(\epsilon^2 p q\right)^{-1} \lesssim n_{\rm d}\ln n_{\rm d}
\label{detachment}
\ee

\begin{figure}[ht]
\epsfig{file=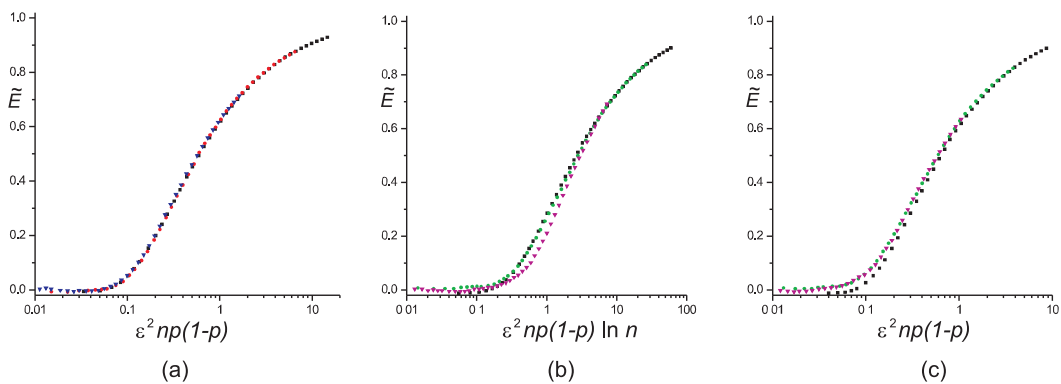, width=14cm} \caption{The dependence of the reduced mean energy $\tilde
{E} = (\la e_n \ra -\la e_0 \ra)/(\la e_{\infty} \ra - \la e_0 \ra)$ of the optimal path on the
reduced size $n$ of the system. (a) for $c=4$ and $\epsilon =  0.3$ (black squares), $\epsilon
=0.2$ (red circles), and $\epsilon =0.1$ (blue triangles); (b) and (c) for $\epsilon = 0.2$ and
$c=2$ (black squares), $c=8$ (green circles), and $c=32$ (magenta triangles). Note that curves for
$c=2,8$ almost collapse after rescaling prescribed by the r.h.s of \eq{detachment}, while those for
$c=8,32$ collapse with rescaling prescribed by the l.h.s. of \eq{detachment}.}
\label{fig:length}
\end{figure}

In \fig{fig:length} we show the results of our computer simulation of the average energy of the
optimal path as a function of the sequence length for different values of $a$ and $p$. One notes
the crossover (for fixed $a$ and $p$) from the path sticking to the diagonal at low $n$ and the
high--$n$ regime, where the optimal path is detached. For $n\gg 1$ the average energy of the path
eventually saturates at some value $E_{\infty}$, which is $a$-- and $p$-- dependent. Moreover,
though the detachment point is not exactly well--defined, the rescaling according to the inequality
\eq{detachment} shows that it gives rather decent estimate of the detachment point. Note also that
the plateau region persists up to quite large values of $\epsilon$. Indeed, it is easy to see from
\eq{eq_a} that the detachment happens at $n_d>2$ (and thus a plateau of at least two points exists)
for any $a> a_d = 1/2$. It is less obvious and more important, however, that the more accurate
estimate \eq{detachment} is still relevant in the whole range of $a\in(1/2,1)$.

\subsection{The general case $a \in [0,1]$: energy cost and fluctuations.}
\label{sect:4c}

\begin{figure}[ht]
\epsfig{file=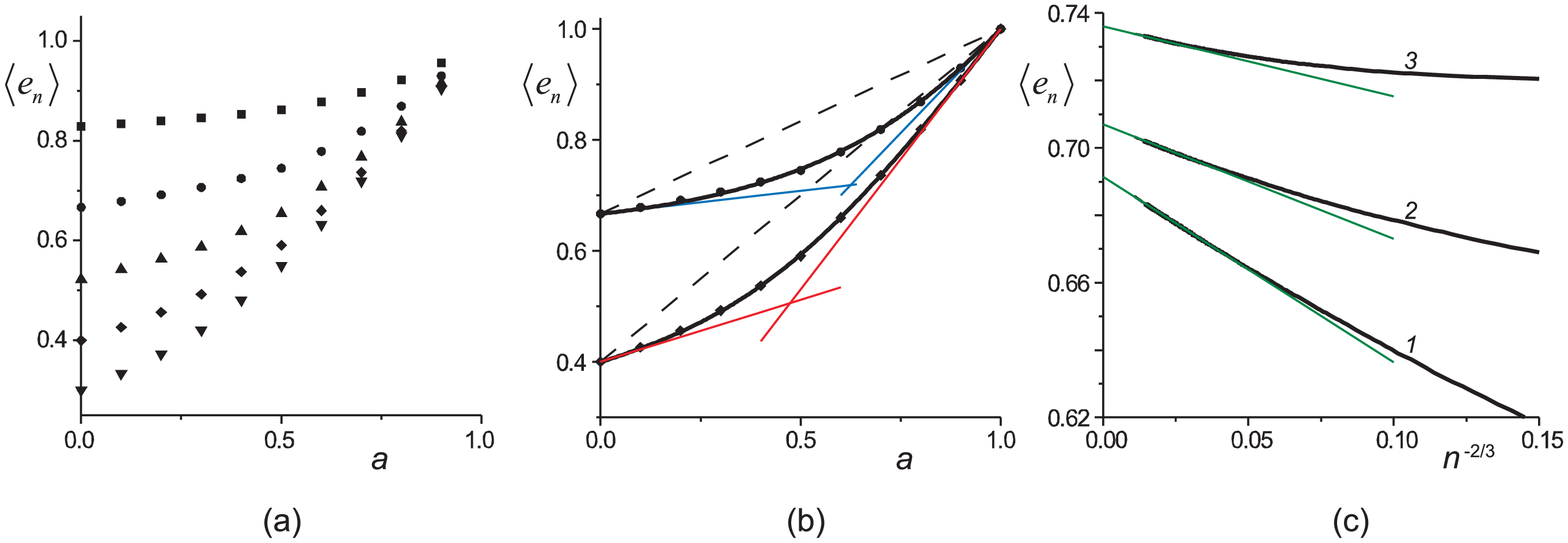, width=13cm} \caption{(a) The limiting value of the ground state energy
per cite $\la e_n(a)\ra \equiv \la E(a)\ra /n$ as a function of $a$ for different $c$: $c = 2$
(squares), $c=4$ (circles), $c=8$ (up triangles), $c = 16$ (diamonds), $c=32$ (down triangles); (b)
The upper (dashed line) and lower (thin solid lines) bounds and the hyperbolic fit (thick line) of
the $\la e_n(a) \equiv \la E(a)\ra /n\ra$ dependence for $c=4$ (circles) and $c=16$ (diamonds); (c)
The examples of the ground state energy per cite $\la e_n \ra$ as a function of $n^{-2/3}$ (thick
line) and the finite--size scaling fits used to obtain points in the figure a) (thin lines) for
several different values of $a$ and $c$, line 1: $a=0.2, c=4$, line 2: $a=0.6, c=8$, line 3:
$a=0.7, c=16$.}
\label{fig:e(a)}
\end{figure}

Consider now the general case of $a \in [0,1]$. In \fig{fig:e(a)} we present the estimates of the
average ground state energy $\la e_n(c,a) \ra = \la E_n(c,a)\ra /n$ for different values of $c$ and
$a$. These estimates we obtain by the finite size scaling extrapolating $\la e_n(c,a) \ra$ from
large, but finite, $n$ to $n\to\infty$.

In our construction we use the following conjecture. One sees from \eq{nech_maj_energy} that at
$a=0$ and for $m=n\gg 1$ the average ground state energy, $\la e_n(c,a=0) \ra$, converges to its
value at infinity, $\la e_{\infty}(c,a=0) \ra=\frac{2}{1+\sqrt{c}}$, with the scaling exponent
$\alpha=-2/3$:
\be
\la e_n(c,a=0) \ra = \frac{1}{n}\la E_{n,n}(c,a=0)\ra = \frac{2}{1+\sqrt{c}} + \frac{c^{1/6}
(\sqrt{c}-1)}{\sqrt{c}+1} \la \chi \ra n^{-2/3} = \la e_{\infty}(c,a=0) \ra + f(c) \la \chi \ra
n^{\alpha}
\label{eq:e}
\ee
where $\la \chi \ra = -1.7711...$ (see \cite{TW}).

We assume that the critical exponent $\alpha$ is $a$--independent and the finite size scaling of
$\la e_n(c,a) \ra$ for $a>0$ and $n\gg 1$ reads (see also \cite{Hwa97})
\be
\la e_n(c,a) \ra = \la e_{\infty}(c,a) \ra + g(c,a) \la \chi \ra n^{\alpha}
\label{eq:e2}
\ee
where $g(c,a)$ is some function of $c$ and $a$, but not of $n$. Extrapolating the data of $\la
e_n(c,a) \ra$ computed numerically for large finite $n$ to $\la e_{\infty}(c,a) \ra$ on the basis
of finite size scaling \eq{eq:e2}, we arrive at the family of curves $\la e_{\infty}(c,a) \ra$ for
$c=2,4,8,16,32,64$ shown in \fig{fig:e(a)}a,b. The results presented in \fig{fig:e(a)}c, as well as
those of \cite{Hwa97} demonstrate that the conjecture \eq{eq:e2} is actually plausible. Apart from
the points obtained by numerical simulation, in \fig{fig:e(a)}b we depict: a) the estimates for
$\la e_{\infty}(c,a) \ra$ at small $a$ given by the inequality \eq{E_ineq}, and b) the estimates of
$\la e_{\infty}(c,a) \ra$ on the plateau for $a\to 1$ (Eq.\eq{E_diag2}).

One should note that the numerical results for $a\ll 1$ are very close to the lower bound of
\eq{E_ineq}. We use this fact to produce a fit for the dependence $\la e_{\infty}(c,a)\ra$ in the
whole range of parameter $a\in [0,1]$ for few values of $c$ ($c=4,16,64$). Namely, we fit the data
of $\la e_{\infty}(a) \ra$ by a hyperbola of general form
\be
(\la e_{\infty}(a) \ra + \kappa_1 a + \delta_1) (\la e_{\infty}(a) \ra + \kappa_2 a + \delta_2) = R
\label{hyperbola}
\ee
with the constraints that this hyperbola passes through the points $(a,e_{\infty}(a))=
(0,2/(\sqrt{c}+1))$ at $a=0$, and $(a,e_{\infty}(a))=(1,1)$ at $a=1$ with the slopes given by
limiting linear approximations \eq{E_ineq} and \eq{E_diag2} correspondingly. These four constraints
leave us effectively with only one free parameter, which we change to arrive at the best fit of the
experimental data. As one sees from \fig{fig:e(a)}b, the found fits for different values of $c$ are
quite good.

Let us now discuss briefly the fluctuations of the average free energy and their dependence on $n$.
One expects for $n \gg 1$ the average fluctuations $\sigma^2_E$ to be proportional to $n^{2/3}$,
typical for the Kardar--Parisi--Zhang universality class \cite{Hwa97}. This conjecture is
consistent with the computation of the fluctuations of the averaged length of the Longest Common
Subsequence (LCS) in the $a=0$ limit for Bernoulli Matching model (see \cite{nech_maj}):
\be
\sigma^2_E(n) = {\rm Var}\, E_{n,n}(c) = \la E_{n,n}^2(c) \ra - \la E_{n,n}(c) \ra^2 \approx
\left(\langle\chi^2\rangle- {\langle\chi\rangle}^2\right)\, f^2(c)\, n^{\theta_0}
\label{fluct}
\ee
where $\theta_0=2/3$ and $\langle \chi^2\rangle-\langle \chi\rangle^2= 0.8132\dots$.

The behavior for intermediate values of $n$ is more involved. In particular, for small $a$ and
intermediate $n$ one expects for $\sigma^2_E(n)$ the growth with the critical exponent $\theta_1$:
\be
\sigma^2_E(n) \sim n^{\theta_1}
\label{growth1}
\ee
The exponent $\theta_1$ is known to be typical for the "transitional" regime in the (1+1)D KPZ
equation \cite{Krug,Krech}. In terms of the work \cite{Krech} the exponent $\theta_1$, which
governs the short--time behavior of the correlation function of KPZ model, is $\theta_1 =
(d+4)/z-2$, where $z$ is the dynamic exponent \cite{Krech}, and $d$ is the space dimensionality. In
$d=1$ the value of $z$ for KPZ model is known exactly, $z=3/2$, giving the value $\theta_1=4/3$.

For $a= 1-\epsilon$ ($\epsilon\ll 1$) the plateau regime for $e_n(c,a)$ exists at low $n\lesssim
n_{\rm d}$ (where $n_{\rm d}$ is defined in \eq{detachment}). The arguments of Section
\ref{sect:4b} allow us to expect the in this case the variance $\sigma^2_E(n)$ behaves as
\be
\sigma^2_E(n) \sim n^{\theta_2}
\label{growth2}
\ee
with the Gaussian exponent $\theta_2=1$ since the plateau energy is just the sum of $n$ independent
random variables.

\begin{figure}[ht]
\epsfig{file=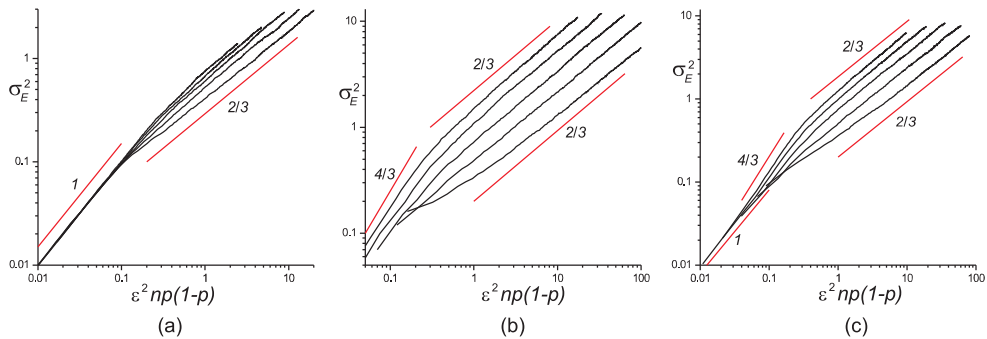,width=13cm} \caption{The dispersion $\sigma_E$ of the ground state energy
as a function of $N$ for different values of $a$ and $c$. (a) $a = 0.7$, (b) $a=0.2$, (c) $a=0.4$.
In all figures $c= 2,4,8,16,32$ in ascending order.}
\label{fig:disp}
\end{figure}

The numerical results presented in \fig{fig:disp} for $\sigma^2_E(n)$ fully confirm the behaviors
\eq{fluct}, \eq{growth1} and \eq{growth2}. In the case of intermediate $a$ shown in \fig{fig:disp}c
the sequence of regimes, at least for large $c$ is more reach: we first note the exponent
$\theta_2=1$ (plateau), then the exponent $\theta_1=4/3$ ("transitional" KPZ), and finally the
exponent $\theta_0=2/3$ (large scale KPZ). It looks like the growing plateau region continuously
"swallows up" the finite--size KPZ region with the increase of $a$, and thus at $\epsilon=1-a \ll
1$ one sees only two regimes.

The important question concerning the results presented above is
how universal are they with respect to the change of the model
settings. Indeed, the exponents $\theta_{0,1,2}$ are obtained in
the assumption of Bernoulli matching, i.e. assuming the contact
matrix $\beta$ doesn't have any correlations in it. As it was
mentioned above, in reality it is not the case. Though we assume
that some of the exponents may be universal, and may even hold in
the case when there are strong long-range correlations in the
structure of the associating DNA strands, the situation is {\it a
priori} very unclear, a thorough numerical investigation of this
question would be, in our opinion, an essential contribution in
the field of the KPZ-related studies.

\section{Conclusion}
\label{sect:5}

In this work we have analyzed the average normalized ground state energy, $e$, of the complex of two
random heteropolymers with quenched sequences as a function of chain length, $n$, in the ensemble
of chains with uniform distribution of primary structures. The main attention is paid to the
behavior of the function $e(n)$ at intermediate chain lengths and low temperatures.

The dependence $\la e_n \ra$ is shown in \fig{fig:length}. Besides the formal estimates of the
boundaries \eq{E_ineq}, \eq{E_diag2}, and of the crossover length, $n_{\rm d}$
(Eq.\eq{detachment}), it seems to be desirable to acquire the qualitative understanding of the
zipping energy $\la e_n \ra$ for different chain lengths and different values of $a$.

One sees that the normalized energy $\la e_n \ra$ for relatively long ($n>n_{\rm d}$) zipped chain
configurations, is larger than the corresponding energy in a hairpin state for $n<n_{\rm d}$. The
reason for this result is as follows. Longer chains could optimize their energy matching via loops creation while
for short chains the penalty for loop formation is forbiddingly large. Hence the inequality
\eq{detachment} gives the criterium for characteristic scale length which separates two kinds of
structure behavior: short chains form the hairpin configuration in which the monomers are forced to
bond without any regard of their species, while long chains are capable of adjusting their spatial
configurations by loop formation to obtain better matching. The crossover around $n_{\rm d}$ is,
thus, separating the small $n$ region where the energy approaches the plateau value \eq{E_diag2}
exponentially fast with decreasing $n$, and infinitely large region of increasing $\la e_n \ra$
where it approaches its value at $n\to\infty$ with the power low dependence $\la e_{\infty} \ra
-\la e_n \ra \sim n^{-2/3}$. This behavior of $\la e_n \ra$ depends  only qualitatively (see
\eq{detachment}) on the parameter $a$ for sufficiently large $a>a_{\rm d}\sim 0.5$.

\begin{figure}[ht]
\epsfig{file=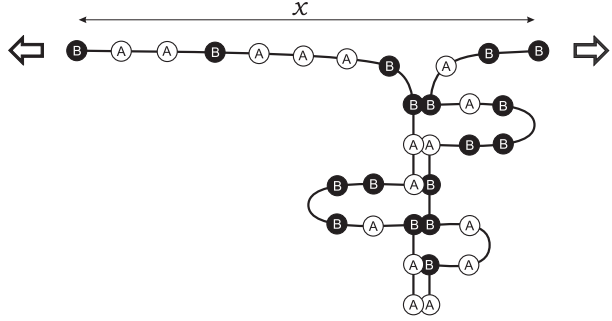,width=8cm} \caption{Unzipping of two random heteropolymers.}
\label{fig:unzip}
\end{figure}

The unzipping process of two random heteropolymer chains is schematically shown in \fig{fig:unzip}.
The results of previous sections allow us to find the dependence of the force $f(x)$ per chain
monomer, on an average extension distance, $x$, between chain ends. If $N$ is the total length of
each heteropolymer chain, and $n$ is the average current length of the heteropolymer complex
measured from its common bottom end (see \fig{fig:unzip}), then by construction, $x=2(N-n)$. For
the sake of simplicity, we neglect here the fluctuations of the unzipped regions of the chain.

\begin{figure}[ht]
\epsfig{file=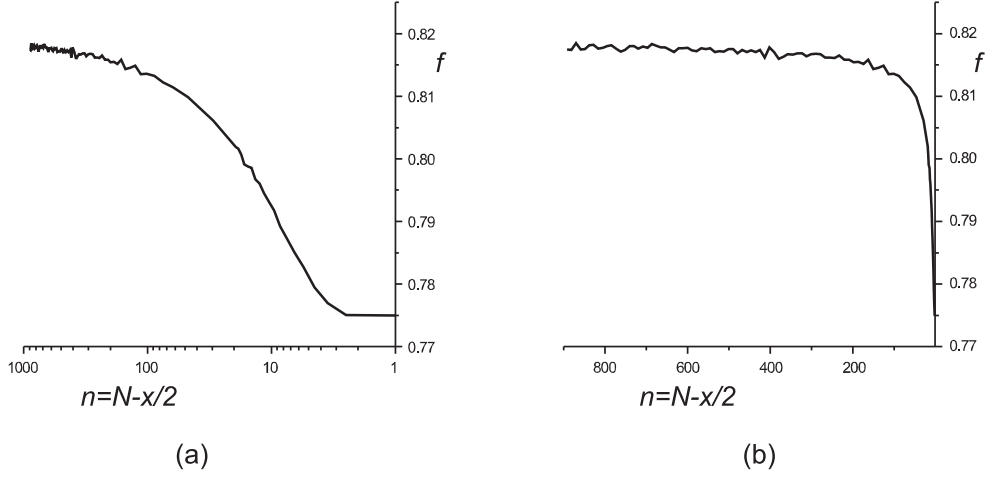,width=13cm} \caption{Dependence of unzipping force, $f$, per chain
monomer on average separation distance, $x$. (a) log-linear scale, (b) linear scale.}
\label{fig:9}
\end{figure}

The plot of the average force, $f$, per chain monomer on the average separation distance, $x$, is
shown in \fig{fig:9}. To be precise, $f(x)$, is the force necessary to unzip two random
heteropolymer chains whose ends are separated by the distance $x$ averaged over all equally
distributed primary structures at low temperatures for fixed value $a=\frac{v}{u}$ and given number
of letters in the alphabet, $c$. The function $f(x)$ can be easily obtained from the dependence
$\la e_n \ra$ shown in \fig{fig:e(a)}. Namely, $f(x) = \frac{d}{dn} (n \la e_n \ra)$ at $n=N-x/2$.

Qualitative explanation of this phenomenon repeats the above discussion of the ground state free
energy $\la e_n \ra$. As it has been said already, the main attention in our work is paid to
relatively small $n$, i.e. large average separation distances, $x$. (For discussions of the
peculiarities of the force on the other bound, i.e. at $x \to 0$, see \cite{singh}.) When $x$
approaches the contour length, $2N$, the equilibrium unzipping force $f(x)$ gradually decreases as
$const-n^{-2/3}=const - (N-x/2)^{-2/3}$ until $N-x/2 \sim n_d$ when the force drops further down to
reach the limiting plateau value \eq{E_diag2} where it saturates independently of further increase
of $x$.

Let us stress once more that the result obtained is valid only for
values of $f(x)$ averaged over the ensemble of realizations of
different heteropolymer sequences: for any given heteropolymer
sequence, the equilibrium force would be a highly fluctuating
function of the distance $x$. In reality, moreover, the unzipping
experiments are often set up in the fixed-force ensemble, instead
of the fixed-distance one (see, for example,
\cite{Cocco1,Cocco2}), i.e. the constant force is applied to the
ends of the chain, and the dynamics of the unzipping under this
constant force is studied. In such a setting the knowledge of the
characteristic occupation times for the intermediate states allows
to reconstruct the overall free energy landscape. We predict that
after the averaging over many realizations of such an experiment
with different primary structures, one expects the typical
occupation times for almost unzipped intermediate states to be
less than those for the almost zipped conformations (the
particular difference depends on the applied force).
Correspondingly, the life time of the intermediate states is
gradually decreasing with the increase of $x$ until saturating at
$N-x/2 \sim n_d$.

\begin{acknowledgments}

We are grateful to S. Majumdar for valuable discussion of matching problem. M.V.T. acknowledges
warm hospitality during the stay at LPTMS where this work was started and completed. The work is
partially supported by the grant ACI-NIM-2004-243 "Nouvelles Interfaces des Math\'ematiques"
(France).

\end{acknowledgments}

\begin{appendix}

\section{Average value of the minimum of two independent increments}
\label{app:a}

First of all we should make a conjecture about the distribution of intervals $d_m=m_{i+1}-m_i$
and $d_n=n_{i+1}-n_i$. It seems to be rather natural to suppose that the intervals $d_{m,n}$ have
the exponential distribution, i.e. $p(d_{m,n}) \sim e^{-k d_{m,n}}$ (one can easily check that
at least the tails of this distribution are indeed exponential). Normalizing $p(d_{m,n})$, we
get
\be
p(d_{m,n})=\frac{e^{-k d_{m,n}}}{\sum\limits_{d_{m,n}=1}^{\infty}e^{-k d_{m,n}}} = (e^{k}-1)e^{-k
d_{m,n}}
\label{distr}
\ee
The mean values $\la d_m \ra $ and $\la d_n \ra$ are
\be
\la d_m \ra = \la d_n \ra = \sum\limits_{d_{m,n}=1}^{\infty}d_{m,n}\; p(d_{m,n}) = \frac{e^{
k}}{e^{k}-1}
\label{mean}
\ee

Now we are to find the averaged joined minimum $\la d_{\min} \ra$ of two random variables $d_m$ and
$d_n$ distributed with \eq{distr}. To do that we proceed as follows. First of all find the discrete
integral distribution function, $F_1(z)$, for each random distribution, $p(d_m)$ and $p(d_n)$:
\be
F_1(z) = \sum\limits_{d_{m,n}=1}^{z}p(d_{m,n})=1-e^{-k z}
\label{int1}
\ee
Following the general procedure, define now the joined discrete integral distribution function,
$F_2(z)$,
\be
F_2(z)=1-(1-F_1(z))^2=1-e^{-2 k z}
\label{int2}
\ee
Taking the discrete derivative, $p_2(z)=F_2(z)-F_2(z-1)$, we find the probability distribution,
$p_2(z=d)$ for the minimum $d_{\min}=\min [m_{i+1}-m_i,\; n_{i+1}-n_i]$. The last step consists
in taking average $\la d_{\min} \ra$ with respect to the joined distribution function $p_2(d)$:
\be
\la d_{\min} \ra = \sum_{z=1}^{\infty} z\, p_2(z) = \frac{e^{2 k}}{e^{2 k}-1}
\label{av_d}
\ee
Collecting \eq{0_dist}, \eq{mean} and  \eq{av_d}, we get
\be
\left\{
\begin{array}{l}
\disp \frac{e^{k}}{e^{k}-1} = \frac{1+\sqrt{c}}{2} \medskip \\
\disp \frac{e^{2 k}}{e^{2 k}-1}=\la d_{\min} \ra
\end{array} \right.
\label{d_c}
\ee
and thus, resolving \eq{d_c},
\be
\la d_{\min} \ra = \frac{(1+\sqrt{c})^2}{4\sqrt{c}}
\label{d}
\ee
Substituting \eq{d} into \eq{0_deltaE} one obtains finally the estimate of $\Delta E$ from below:
\be
\Delta E>\la L_{n,n}\ra \Big(\la d_{\min} \ra - 1 \Big)a =
\left(\frac{1+\sqrt{c}}{2\sqrt{c}}-\frac{2}{1+\sqrt{c}}\right)na
\label{E_min}
\ee

\section{Auxiliary construction for estimation of the detachment length}
\label{app:b}

Assuming $\frac{n}{c} \gg 1$ one can replace the binomial distribution \eq{binom} with the Gaussian
one and approximate $W(k)$ as follows
\be
\tilde{W}(k,n)=\frac{1}{\sqrt{2 \pi npq}} \exp\left(-\frac{(k-\la k \ra)^2}{2 npq}\right)
\label{binom_ap}
\ee
where $\la k \ra = n q$. The distribution function for the variable $k'$ $W(k',n-1)$ is completely
similar but the replacement $n\to n-1$.

We are now to compute the mean minimal value $\la k'_{N_{\rm ind}} \ra$ of $N_{\rm
ind}$ random variables, each distributed with $W(k',n-1)\equiv W(k')$. Repeating the same procedure
as in the Appendix \ref{app:a}, we proceed as follows. First of all pass to the integral
distribution function, $\tilde{F}(z)$:
\be
\tilde{F}(z) = \int_{-\infty}^z \tilde{W}(k') dk' = \frac{1}{2} \left(1+{\rm
erf}\left[\frac{z-(n-1)q}{\sqrt{2 (n-1)pq}}\right] \right)
\label{int_f}
\ee
Now construct the new probability distribution function, $Q(z)$, for the joint distribution, as
follows:
\be
Q(z) = \frac{d}{dz}\left[1-(1-\tilde{F}(z))^{N_{\rm ind}}\right]= N_{\rm ind}\, \tilde{F}'(z)\,
(1-\tilde{F}(z))^{N_{\rm ind}-1}
\label{q}
\ee
The desired mean minimal value $\la k'_{N_{\rm ind}} \ra$ reads now
\be
\la k'_{N_{\rm ind}} \ra = \int_{-\infty}^{\infty} z Q(z) dz
\label{mean_k}
\ee
Now, taking the estimate \eq{ind} into account one readily arrives to the lower bound for $\la k'_{N_{\rm ind}}
\ra$. Indeed, for $N_{\rm ind}=2$
\be
\la k_{N_{\rm ind}}^{'\min} \ra = \frac{1}{\sqrt{\pi}} \int_{-\infty}^{\infty}
\left(y\sqrt{2(n-1)pq} + (n-1)q\right) e^{-y^2} \Big(1-{\rm erf}(y)\Big) dy = (n-1)q -
\frac{1}{\sqrt{\pi}}\Big((n-1)pq\Big)^{1/2}
\label{mean_k2}
\ee
For $N_{\rm ind}=3n-2\gg 1$ (see \eq{ind}) the integral \eq{mean_k} cannot be computed analytically
and therefore one needs to apply some approximative approach. We proceed as
follows. The function $\tilde{F}(z)$ has a sense of the area under the curve $\tilde{W}(k)$ in the
interval $k'\in (-\infty,z]$. Consider now $N_{\rm ind} \gg 1$ independent random variables each
distributed with $\tilde{W}(k')$. For $\frac{z-(n-1)q}{\sqrt{(n-1)pq}} \ll -1$
on average one point of $N_{\rm ind}$ equally distributed random points lies in the area
$\tilde{F}(z) \sim N_{\rm ind}^{-1}$. Since this area is the area under the left tail of the
distribution $\tilde{W}(k')$, the point inside this area is the minimal one by construction. So,
expanding $\tilde{F}(z)$ for $\frac{z-(n-1)q}{\sqrt{(n-1)pq}} \ll -1$, we get
\be
\tilde{F}(z) = \frac{1}{2}\left(1+{\rm erf}\left[\frac{z-(n-1)q}{\sqrt{2(n-1)pq}}\right]\right)
\simeq \frac{\sqrt{(n-1)pq}}{\sqrt{2\pi}((n-1)q-z)}\exp\left(-\frac{(z-(n-1)q)^2}{2(n-1)pq} \right)
\sim \frac{1}{N_{\rm ind}}
\label{tail}
\ee
Since the term in the exponent in \eq{tail} varies much faster than the pre-exponential term, we
can roughly estimate $z=\la k_{N_{\rm ind}}^{'\max} \ra$ as follows
\be
\la k_{N_{\rm ind}}^{'\max} \ra \simeq (n-1)q - \Big(2(n-1)pq\Big)^{1/2} \left[\ln \left(N_{\rm
ind} \Big((n-1)pq\Big)^{1/2} \right)\right]^{1/2}
\label{mean_z}
\ee
Note that Eq.\eq{mean_z} is obtained from Eq.\eq{tail} under the condition $z<(n-1)q$
which fixes the right sign of the square root branch of the second term in Eq.\eq{mean_z}.

Substituting $N_{\rm ind}=3n-2$ into \eq{mean_z} and taking into account that $n\gg 1$, we get the
following desired estimate for $\la k_{N_{\rm ind}}^{'\max} \ra$:
\be
\la k_{N_{\rm ind}}^{'\max} \ra \simeq (n-1)q - (2npq)^{1/2} \left[\ln \left(n^{3/2}(pq)^{1/2}
\right)\right]^{1/2}
\label{mean_z2}
\ee
Now we can use the boundaries \eq{mean_k2} and \eq{mean_z2} for getting lower and upper bounds of
$\la \delta k_{N_{\rm ind}} \ra$ and of the detachment length, $n_{\rm d}$ -- see
Eqs.\eq{deltaK}--\eq{detachment}.

\end{appendix}

\end{document}